\documentclass[conference]{IEEEtran}
\IEEEoverridecommandlockouts
\usepackage{cite}
\usepackage{amsmath,amssymb,amsfonts}
\usepackage{algorithmic}
\usepackage{graphicx}
\usepackage{textcomp}

\usepackage{float}
\usepackage{textcomp}
\usepackage{setspace}
\usepackage{psfrag}
\usepackage{color,soul}
\usepackage{algorithm}

\def\BibTeX{{\rm B\kern-.05em{\sc i\kern-.025em b}\kern-.08em
    T\kern-.1667em\lower.7ex\hbox{E}\kern-.125emX}}

\begin{document}

\title{Computing by nowhere increasing complexity
\thanks{This research is supported by Israel Science Foundation Grant
No. 1723/16.}
}

\author{\IEEEauthorblockN{1\textsuperscript{st} Bar Y.\ Peled}
\IEEEauthorblockA{\textit{Department of Mechanical Engineering} \\
\textit{Ben-Gurion University of the Negev}\\
Beer Sheva, Israel \\
barp@post.bgu.ac.il}
\and
\IEEEauthorblockN{2\textsuperscript{nd} Vikas K.\ Mishra}
\IEEEauthorblockA{\textit{Department of Mechanical Engineering} \\
\textit{Ben-Gurion University of the Negev}\\
Beer Sheva, Israel \\
vikas.mishra45@gmail.com}
\and
\IEEEauthorblockN{3\textsuperscript{rd} Avishy Y.\ Carmi}
\IEEEauthorblockA{\textit{Department of Mechanical Engineering} \\
\textit{Ben-Gurion University of the Negev}\\
Beer Sheva, Israel \\
avcarmi@bgu.ac.il}
}

\maketitle

\begin{abstract}
	A cellular automaton is presented whose governing rule is that
        the Kolmogorov complexity of a cell's neighborhood may not
        increase when the cell's present value is substituted for its
        future value. Using an approximation of this two-dimensional Kolmogorov complexity the underlying automaton is shown to be
        capable of simulating logic circuits. It is also shown to capture trianry logic described by a
	quandle, a non-associative algebraic structure. A similar automaton
	whose rule permits at times the increase of a cell's neighborhood
	complexity is shown to produce animated entities which can be used as
	information carriers akin to gliders in Conway's game of
        life.
\end{abstract}


\section{Introduction}\label{sec:Introduction}


In natural and artificial systems pattern formation and order are
the imprints of a computational process. We ask whether this process can be
reversed -- can the enforcement of order or a pattern bring about
computation ? What sort of computation can be realized, for example, by
not allowing the local patterns in cellular automata to get any more
complex than they already are ? Can any computation be realized in
such a manner ? This work is an attempt to answer some of these
questions.

A cellular automaton is a discrete dynamical system that was
originally conceived in the late 1940's by John von Neumann and
Stanislaw Ulam. They incorporated Cellular automaton model into von
Neumann's idea of a “universal constructor”
\cite{von1966theory}. Cellular automaton exhibits a new way of
thinking of how little complexity can achieve interesting behavior,
which can be emulated in diverse physical and biological phenomena
\cite{Wolfram2002}. They have played a significant role in computation. As
has been seen that using Conway's game of life, for example, they can
realize a Turing machine. The idea that the universe in itself is a
cellular automaton inspired Zuse's Calculating Space
\cite{zuse1969rechnender}, a precursor of digital physics.

A typical cellular automaton consists of a grid of cells each storing a value and a
set of rules according to which their values change.
Although the underlying rules may be simple, the behavior of
the automaton as a whole may quickly become complex, a trait that
allows such automatons to emulate diverse physical and biological
phenomena. There are many known cellular automatons whose chaotic
behavior has been intensively studied. Perhaps the most famous are the
game of life (Life), and rule 110. It has been shown that both of them can
realize any computation an ordinary computer can perform
\cite{cook2004universality}.

Von Neumann's universal constructor used thousands of cells
and 29 states to self-reproduce. Its complexity
was later on reduced in \cite{codd1968cellular}. Langton constructed a
self-replicating automaton known as Langton's loops
\cite{langton1984self}. His model is less complex than that of the
previous models due to von Neumann and Codd. Langton's model was
further simplified in \cite{byl1989self, reggia1993simple}. The constructions of the
smallest computationally universal cellular automata have been
considered in \cite{lindgren1990universal}. Reversible cellular
automaton has been studied in \cite{margolus1984physics}.
The links between dynamical and computational properties of cellular
automaton have been investigated by Di Lena and Margara
\cite{di2008computational}.


Here we show that computation in cellular automata can be realized by
exerting local order, where ``order'' is mathematically defined in terms of
Kolmogorov complexity. In particular, we formulate a cellular
automaton that employs the following rule. A cell's value is changed
from $y$ to $x$ if the Kolmogorov complexity of its present Moore
neighborhood is smaller with $x$ than with $y$. Using an approximation
of this two-dimensional Kolmogorov complexity we show that the underlying automaton can compute Boolean
functions. In particular, it can realize a universal set of Boolean gates, AND and NOT, as well as wire
elements for transferring information from a given
cell to any other cell. This automaton is also shown to capture
trinary logic described by a quandle, a non-associative algebraic
structure whose axioms have been interpreted as laws of
conservation of information
\cite{carmi2015computing,moskovich2015tangle}. 

The expressiveness of a cellular automaton employing the
``nowhere increasing'' complexity rule is rather limited. The initial
grid which encodes the logic circuit gradually obliterates during the automaton
evolution. This implies that similar constructions may not be used for
recursive computations. Nevertheless, a similar automaton with an
alternating rule where the cell's neighborhood complexity may at times
increase is shown to produce animated entities which can be used as
information carriers akin to gliders in the game of
life~\cite{lizier2008local}.




%

This work is organized as follows. The next section is a brief
introduction to the notion of Kolmogorov complexity. The underlying
cellular automaton is described in Section~\ref{sec:CA}. It is shown
to realize logic gates in Section~\ref{sec:comps}, and other
computations and gliders in Section~\ref{sec:other}. Concluding
remarks are offered in the last section.

\section{Kolmogorov complexity}\label{sec:Others}

The Kolmogorov complexity, $K(s)$,
of an object or a string $s$, is a measure of the computational
resources that are needed to generate $s$
\cite{kolmogorov1965three,chaitin1969length}. Informally, $K(s)$
is the length of the shortest program that produces $s$ as an output
and then halts. The string $\left(01\right)^{n}$, for example, can be
described as ``n repetitions of $01$''. This string is $2n$ digits long while its description
contains around $\log_{2}\left(n\right)$ binary digits, i.e., the
length of the binary representation of $n$. On the other hand, 
seemingly random occurrences of zeros and ones will generally not
admit a description shorter than its own length.

The Kolmogorov complexity can be used to define the measure
\begin{equation}
\label{eq:algprob}
2^{-K(s)}.
\end{equation}
If $K(s)$ is the length of a program to a universal prefix Turning
machine that produces $s$ and then halts, then by Kraft's inequality
\eqref{eq:algprob} may be interpreted as an unnormalized measure of
probability over all such programs~\cite{chaitin1969length}.


Kolmogorov complexity is an incomputable function -- there is no
program which takes a string $s$ as an input and produces the number
$K(s)$ \cite{chaitin1969length}. Normally, $K(s)$ is approximated
using known compression techniques, see e.g.,
\cite{rivals1996compression}. But \eqref{eq:algprob} can also be used
to estimate $K(s)$, assuming one can simulate a large number of Turing
machines that produce $s$. Some of these
machines will halt and some won't. Counting the number of them that
produced $s$ and then halted gives a number $m(s)$. The Kolmogorov
complexity is then approximated, up to a constant, as $-\log_2
m(s)$~\cite{delahaye2012numerical, zenil2015two}.

In this work we are interested in the Kolmogorov complexity of a
two-dimensional object, a $3 \times 3$
binary matrix. As explained in the next section such a matrix
describes the Moore neighborhood of a cell.
The complexity of all these 512 binary matrices have been
recently approximated in \cite{zenil2015two}. Simulating a
large number of two-dimensional Turing machines the probability
\eqref{eq:algprob} was approximated for any given matrix
configuration. An estimate of $K$ was then obtained as described
above. These approximations can be found at
\verb!\https://github.com/algorithmicnaturelab/!
\verb!OACC/blob/master/data/K-3x3.csv!



\section{The cellular automaton}\label{sec:CA}

A cellular automaton is defined by a finite number of ``colored'' cells
together with a set of rules that specify how to manipulate their
colors. The cells of a standard automaton contain binary values,
either ``0'' (black) or ``1'' (white). The purpose of the rules is to
determine the state (color) of a cell at time $t+1$ based on the
values of its neighbors, the set of cells in its vicinity at time $t$.
The automaton evolves by using the rules to determine the next value
for each cell in the grid at time $t$. The new cell values thus
obtained make the grid at time $t+1$.

Let us denote $c_{ij}(t) \in \{0,1\}$, the value at time $t$ of the
cell whose coordinates are $\left(i,j\right)$. At each time step a cell
updates its value according to the following rule.\\[0.5ex]

\noindent

{\bf Rule:} \emph{A cell's value is changed
from $c$ to $1-c$ if the Kolmogorov complexity of its present Moore
neighborhood is smaller with $1-c$ than with $c$.}\\[0.5ex]

This rule is mathematically expressed by:
\begin{equation}
\label{eq:liferule}
c_{ij}(t+1) = \left \{
\begin{array}{ll}
c_{ij}(t), & K_{ij}(t) \leq K'_{ij}(t) \\
1 - c_{ij}(t), & \text{otherwise}.
\end{array} \right.
\end{equation}
Here, $K_{ij}(t)$ is the Kolmogorov complexity at time $t$ of the
Moore neighborhood of $c_{ij}(t)$, a $3 \times 3$ pattern composed of
the cell at location $(i,j)$ together with 8 other cells that surround
it. The Kolmogorov complexity at time $t$ of the same pattern in which
the cell at $(i,j)$ is flipped is represented by $K'_{ij}(t)$. For
example, $K$ and $K'$ may be evaluated for the following pair:
\[
\includegraphics[width=0.2\textwidth]{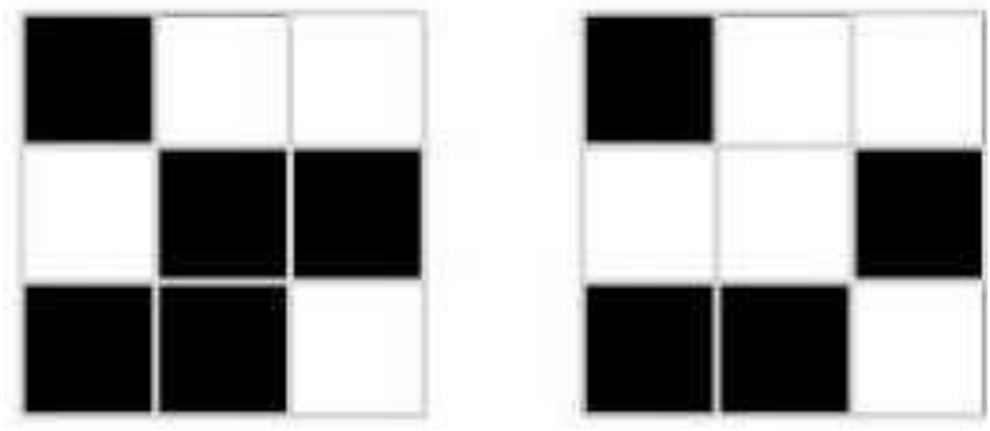}
\]
A pseudocode for a cellular automaton employing the above rule is
provided in Algorithm~\ref{alg:the_alg1}.

\begin{algorithm}
	\caption{Nowhere increasing Kolmogorov complexity CA}
	\label{alg:the_alg1}
	\begin{algorithmic}
		\STATE Syntax: $c_{ij}(t+1) = \text{\bf CA}\downarrow(c_{ij}(t))$
		\STATE Input: $c_{ij}(t)$, $i=1,\ldots, N$, $j=1,\ldots, M$ (grid at
		time $t$)
		\STATE Output: $c_{ij}(t+1)$, $i=1,\ldots, N$, $j=1,\ldots, M$ (grid
		at time $t+1$)
		\FOR{$i = 2:N-1$}
		\FOR{$j = 2:M-1$}
		\STATE Let $A$ be the Moore neighborhood of $c_{ij}(t)$.
		\STATE Obtain $K_{ij}(t)$ using $A$ from the lookup table \cite{zenil2015two}.
		\STATE Flip the value of the middle cell in $A$ and similarly obtain
		$K'_{ij}(t)$.
		\IF{$K_{ij}(t) \leq K'_{ij}(t)$}
		\STATE $c_{ij}(t+1) = c_{ij}(t)$
		\ELSE
		\STATE $c_{ij}(t+1) = 1 - c_{ij}(t)$
		\ENDIF
		\ENDFOR
		\ENDFOR
	\end{algorithmic}
\end{algorithm}


\section{Computational capabilities}\label{sec:comps}

In what follows the underlying automaton is shown to realize the
universal set of gates, NOT and AND, together with wire elements for
connecting them. By a gate or a wire we mean a grid whose cells store
initial values some of which represent inputs and some other represent
outputs. Iterating the above automaton rule where $K$ and $K'$ are
approximated as in \cite{zenil2015two}, changes the cells values. This
procedure is reiterated until all cell values no longer change or
oscillate indefinitely. The output cells then store the outcome of the
computation.

\begin{figure}[htb]
	\psfrag{a}[c]{\color{yellow} $x$}
	\psfrag{b}[c]{\color{yellow} $y$}
	\includegraphics[width=0.5\textwidth]{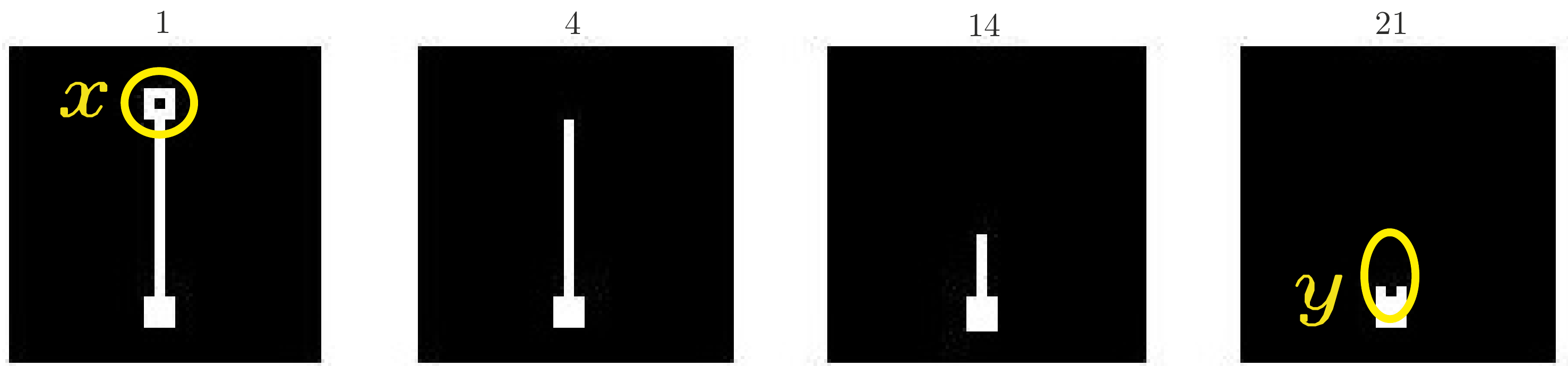} \\
	\includegraphics[width=0.5\textwidth]{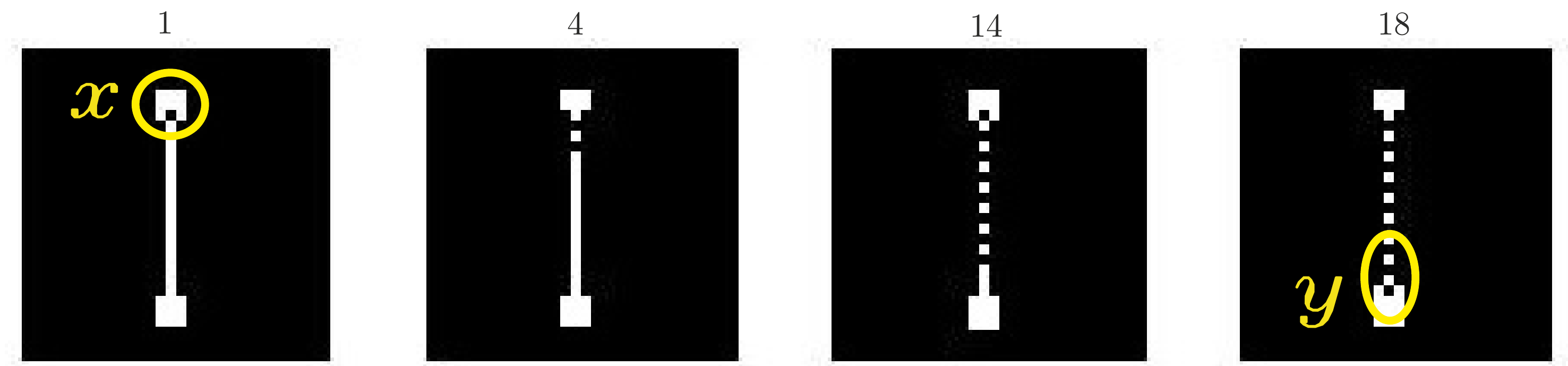}
	
	\caption{Wire element. The grid size is $30\times30$.}
	\label{wire}
\end{figure}

A \emph{wire element} transfers information between cells in the grid.
Its basic form is shown in Figure~\ref{wire}. In this picture the
input $x \in \{0,1\}$ to the wire is specified by a single black cell
in the block of white cells just above the wire upper end. The wire's
other end is connected to another block of white cells. The initial
grids of the automaton for two different inputs $x$ are shown in the
leftmost column in this picture. In the upper left frame the black
cell in the center of the white block represents the input
``0''. Similarly, the input ``1'' in the lower left frame is
represented by a black cell just below the center of this white
block. Injecting ``0'' to the wire completely destructs
it, as seen by the upper sequence of shots taken at different times
during the evolution. Injecting ``1'', the shots in the lower row
show a propagating sequence of alternating black and white cells comes
out of the upper white block all the way down. These two behaviors are
interpreted as a wire carrying either ``0'' or ``1'', i.e., the
content of the wire can be read off at the vicinity of $y$.

\begin{figure}[htb]
	\centering
	\psfrag{a}[c]{\color{yellow} $x$}
	\psfrag{b}[c]{\color{yellow} $y$}
	\includegraphics[width=0.25\textwidth]{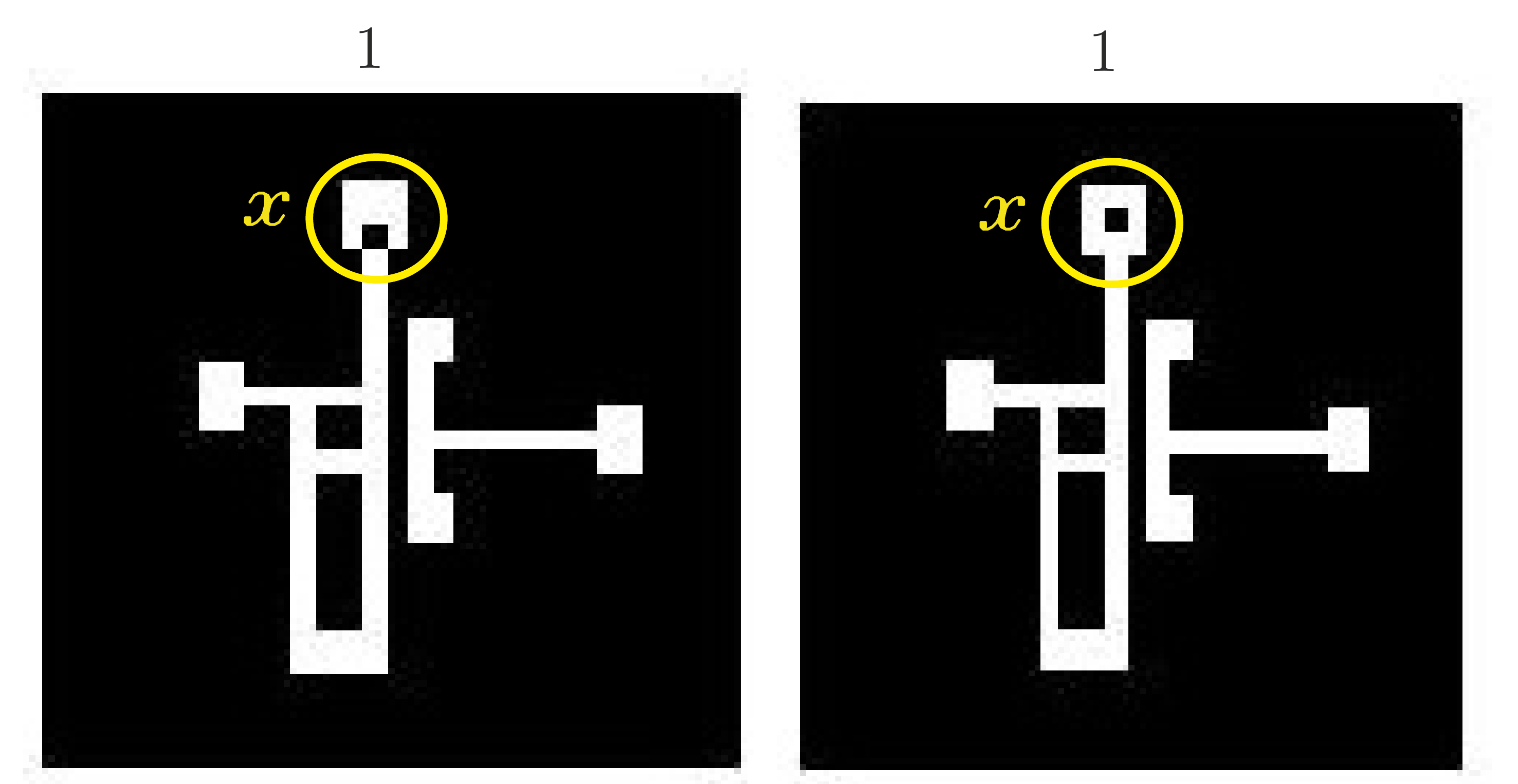} \\
	\includegraphics[width=0.25\textwidth]{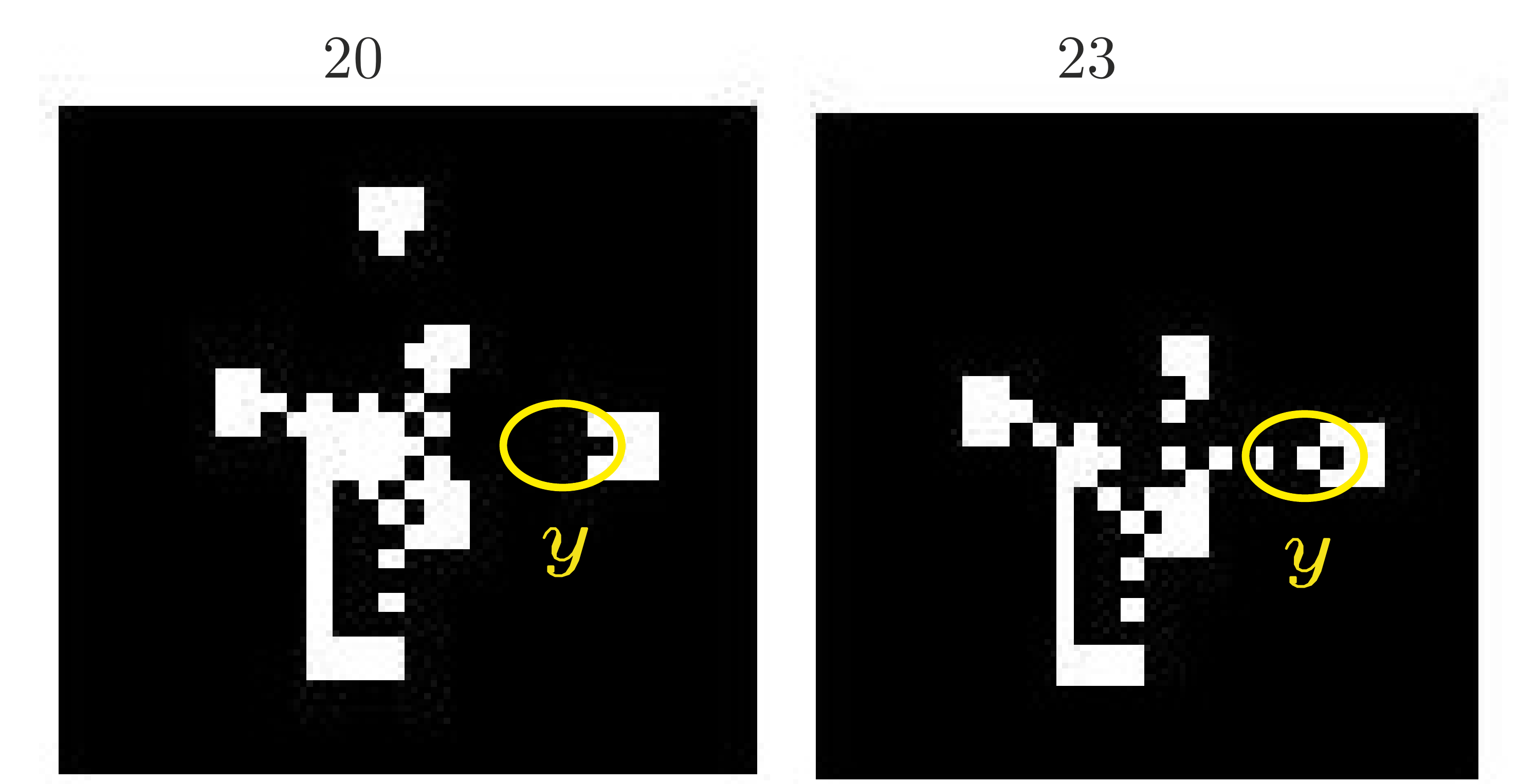}
	\caption{NOT gate. The upper row shows the initial grid for the two
		inputs, $x=1$ (left) and $x=0$ (right). The respective final grid
		for each input is shown in the lower row. The grid size is $30\times30$.} 
	\label{not}
\end{figure}

A \emph{NOT gate} takes inputs $x$ and returns $y=1-x$. Its realization is shown in Figure~\ref{not}.
The upper row in this picture shows the initial grid for this gate
with different inputs, $x=1$ and $x=0$. The respective outputs $y=0$
and $y=1$ in the lower row are obtained after several iterations
of the automaton rule.

\begin{figure}[htb]
	\centering
	\psfrag{a}[c]{\color{yellow} $x$}
	\psfrag{b}[c]{\color{yellow} $y$}
	\psfrag{c}[c]{\color{yellow} $z$}
	\includegraphics[width=0.5\textwidth]{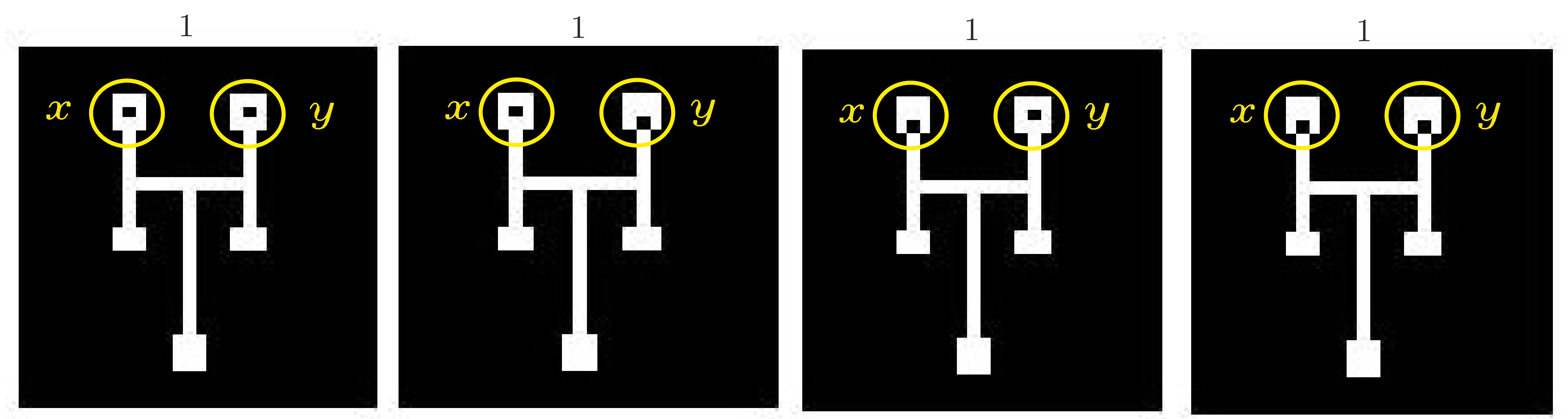} \\
	\includegraphics[width=0.5\textwidth]{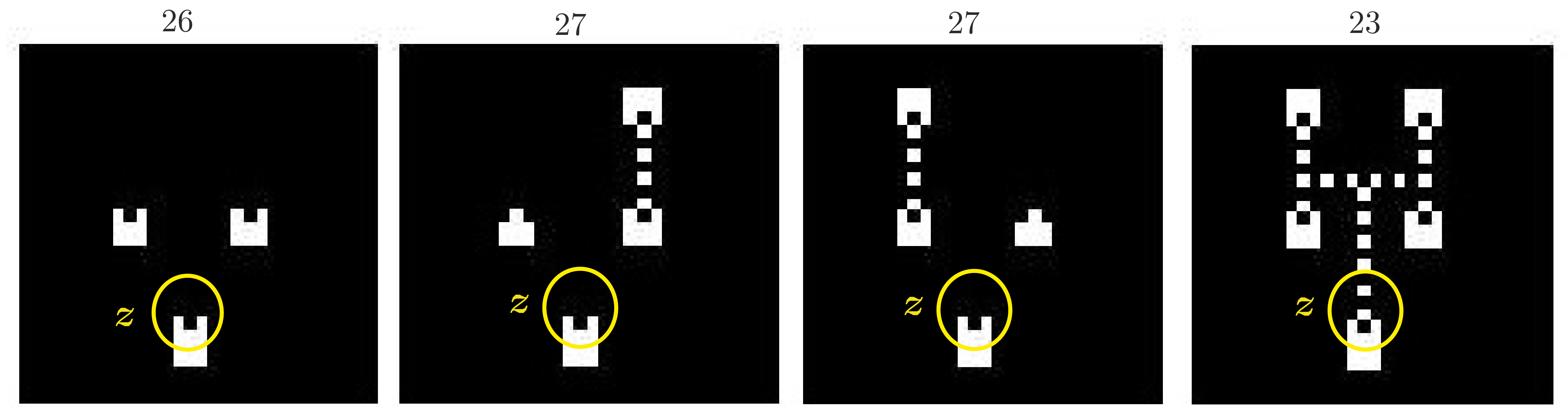}
	\caption{AND gate. The upper row shows the initial grid for the four
		input combinations, from left to right $(0,0)$, $(0,1)$, $(1,0)$, and
		$(1,1)$. The respective final grid for each input is shown in the lower row. The grid size is $30\times30$. }
	\label{and}
\end{figure}

An \emph{AND gate} takes inputs $x$ and $y$ and returns $z=xy$, i.e., only
when $x=y=1$ does the gate returns $z=1$. Its realization is shown in
Figure~\ref{and}. The upper row in this picture shows the initial grid for this gate
with different inputs $(x,y)$, i.e., $(0,0)$, $(0,1)$, $(1,0)$, and
$(1,1)$. The respective outputs $z=0$, $z=0$, $z=0$, and $z=1$ in the
lower row are obtained after several iterations of the automaton rule.

\begin{figure}[htb]
	\centering
	\psfrag{a}[c]{\color{yellow} $x$}
	\psfrag{b}[c]{\color{yellow} $y$}
	\psfrag{c}[c]{\color{yellow} $z$}
	\includegraphics[width=0.5\textwidth]{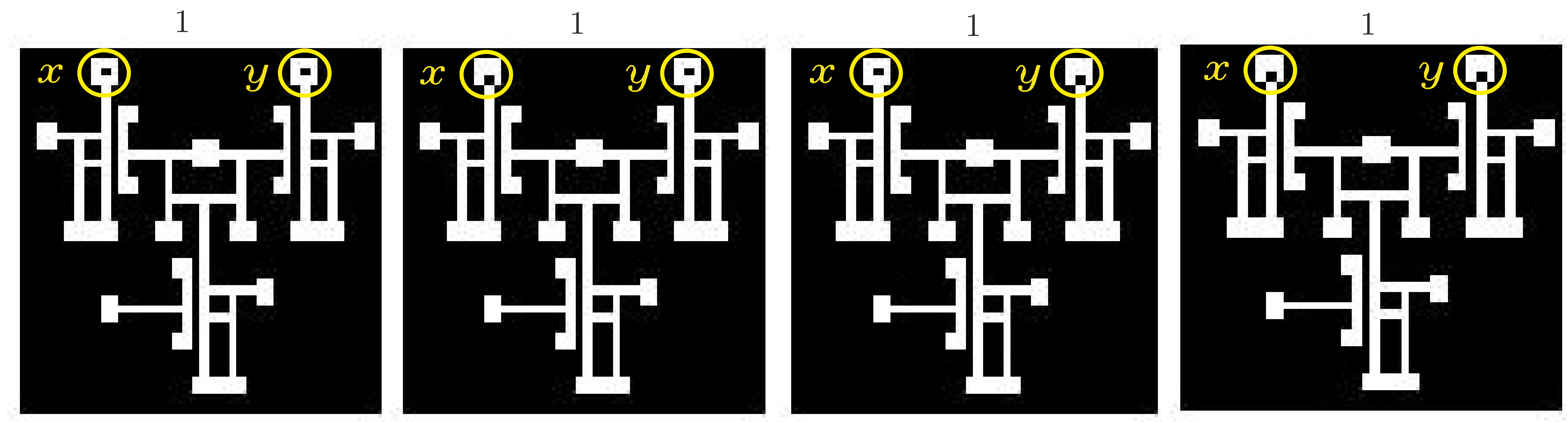} \\
	\includegraphics[width=0.5\textwidth]{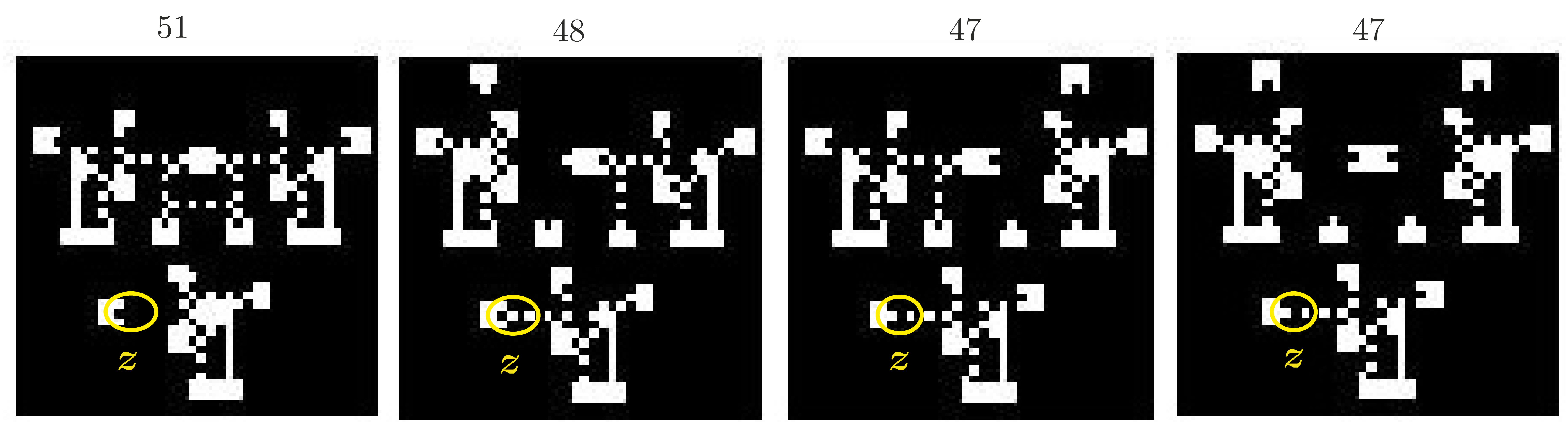}
	\caption{OR gate. The upper row shows the initial grid for the four
		input combinations, from left to right $(0,0)$, $(1,0)$, $(0,1)$, and
		$(1,1)$. The respective final grid for each input is shown in the lower row. The grid size is $40\times40$.}
	\label{or}
\end{figure}

\begin{figure}[htb]
	\centering
	\psfrag{a}[c]{\color{yellow} $x$}
	\psfrag{b}[c]{\color{yellow} $y$}
	\psfrag{c}[c]{\color{yellow} $z$}
	\includegraphics[width=0.5\textwidth]{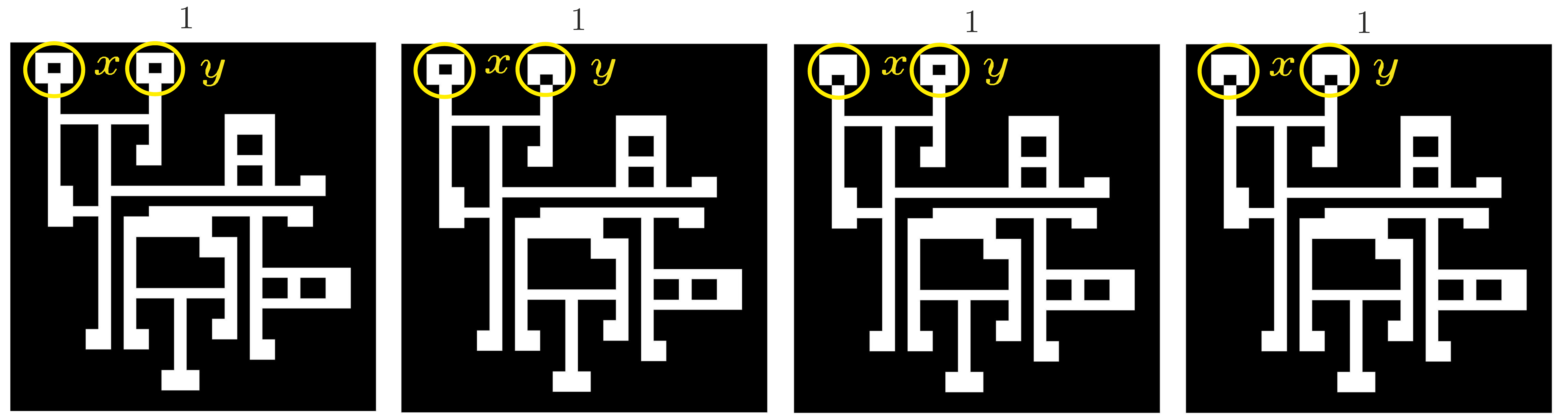} \\
	\includegraphics[width=0.5\textwidth]{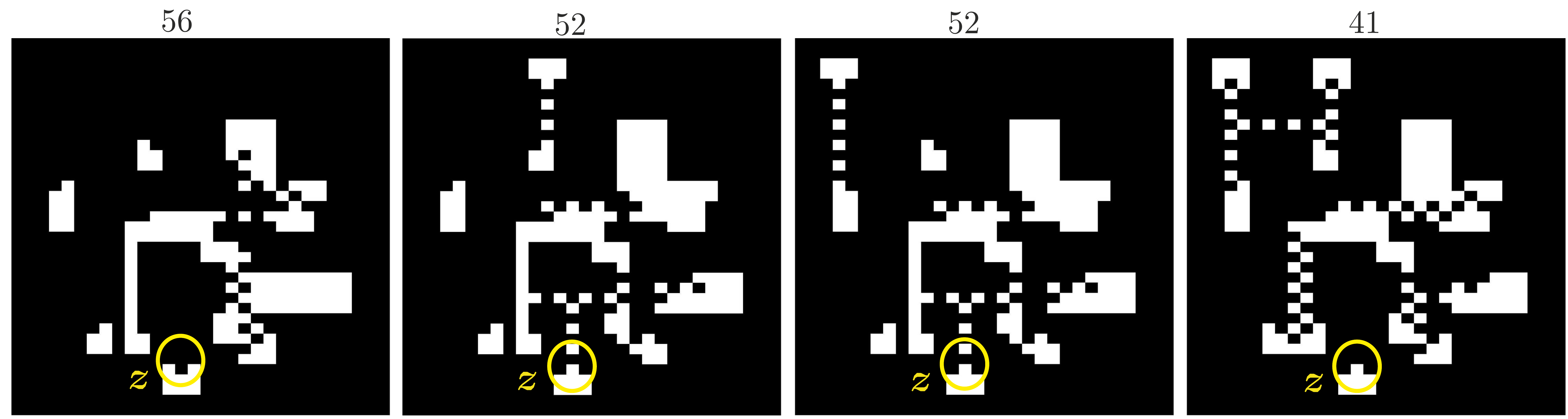}
	\caption{XOR gate. The upper row shows the initial grid for the four
		input combinations, from left to right $(0,0)$, $(0,1)$, $(1,0)$, and
		$(1,1)$. The respective final grid for each input is shown in the lower row. The grid size is $37\times30$.}
	\label{xor}
\end{figure}

An OR gate may be constructed out of an AND and three NOTs, i.e., $z =
1 - (1-x)(1-y)$, so that $z=1$ if at least one of the inputs, $x=1$ or $y=1$. The
realization of this gate is shown in Figure~\ref{or}.
The upper row in this picture shows the initial grid for this gate
with different inputs $(x,y)$, i.e., $(0,0)$, $(1,0)$, $(0,1)$, and
$(1,1)$. The respective outputs $z=0$, $z=1$, $z=1$, and $z=1$ in the
lower row are obtained after several iterations of the automaton rule.
The realization in Figure~\ref{xor} is that of an XOR gate, for which the
output is $z = (1 + (-1)^{xy})/2$.

\section{Gliders and other computations}
\label{sec:other}

Not only binary logic but also trinary logic can be simulated using
the underlying automaton. Consider the following trinary
2-input-2-output gate. Its inputs $x$ and $y$ each may take the values $0$, $1$
or $2$. One of its outputs $z$ is given by $x \triangleright y := (2y
- x) \mod 3$, and the other one is equal $y$, where $\mod$ denotes the
modulo operation. Thus, if the two inputs are not the same then $z$ is
different from both of them, and if the two inputs are the same then
$x$, $y$, and $z$ all are equal.

The operation $\triangleright$ together with the set $Q=\{0,1,2\}$
induce a \emph{quandle} algebra~\cite{carmi2015computing,moskovich2015tangle}. Consider a set
$Q$ together with a set $B$ of binary operations from $Q \times Q$ to
$Q$ with the following properties:
\begin{enumerate}
	\item $x \triangleright x = x$ for all $x \in Q$ and for all $\triangleright \in B$;
	\item $(x \triangleright y)\triangleleft y = x$ for all $x, y \in Q$ and for all $\triangleright, ~ \triangleleft \in B$;
	\item $(x \triangleright y)\triangleright z = (x \triangleright
	z)\triangleright (y \triangleright z)$ for all $x, y, z \in Q$ and
	for all $\triangleright \in B$.
\end{enumerate}
We call the pair $(Q, B)$ satisfying the above properties a $B$-
family of quandles or just a quandle. The above gate satisfies all the
axioms of a quandle, where in this particular case the operations
$\triangleright$ and $\triangleleft$ coincide.

In low-dimensional topology any quandle axiom corresponds
to one Reidemeister move. A sequence of Reidemeister moves relate any
two planar diagrams of the same knot or a tangle. When
these tangle diagrams are colored by pieces of information they act
much like circuits. Here, the instance of a line
cutting through another line, known as a crossing, is seen as a
2-input-2-output gate whose operation is $\triangleright$. As shown in
\cite{carmi2015computing,moskovich2015tangle} the
quandle axioms may then be interpreted as laws of conservation of
information. Indeed, the gates themselves are reversible and their
inputs can be uniquely recovered from the respective outputs.

%

\begin{figure}[htb]
	\centering
	\psfrag{a}[c]{\color{yellow} $x$}
	\psfrag{b}[c]{\color{yellow} $y$}
	\includegraphics[width=0.2\textwidth]{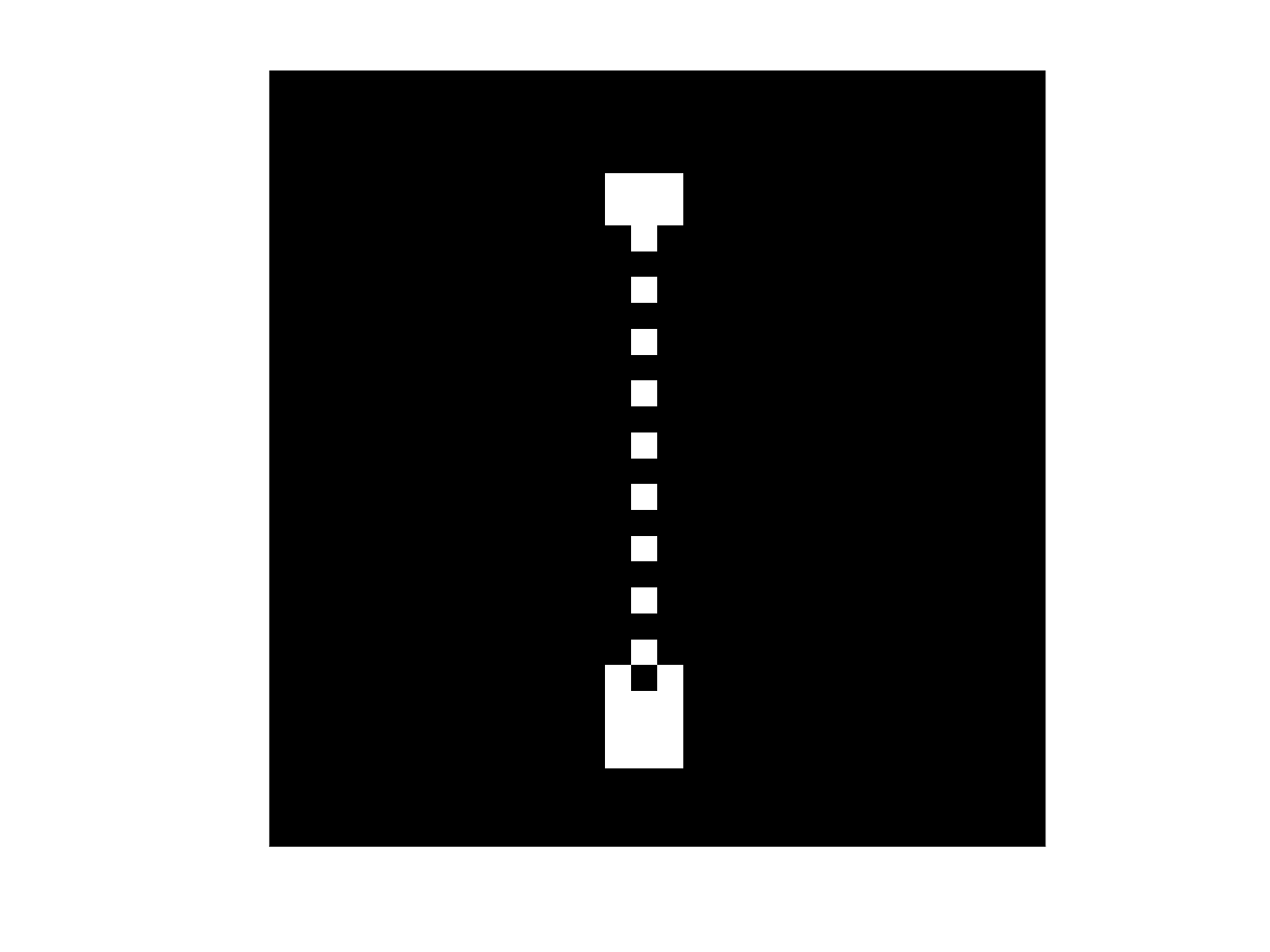}
	\includegraphics[width=0.2\textwidth]{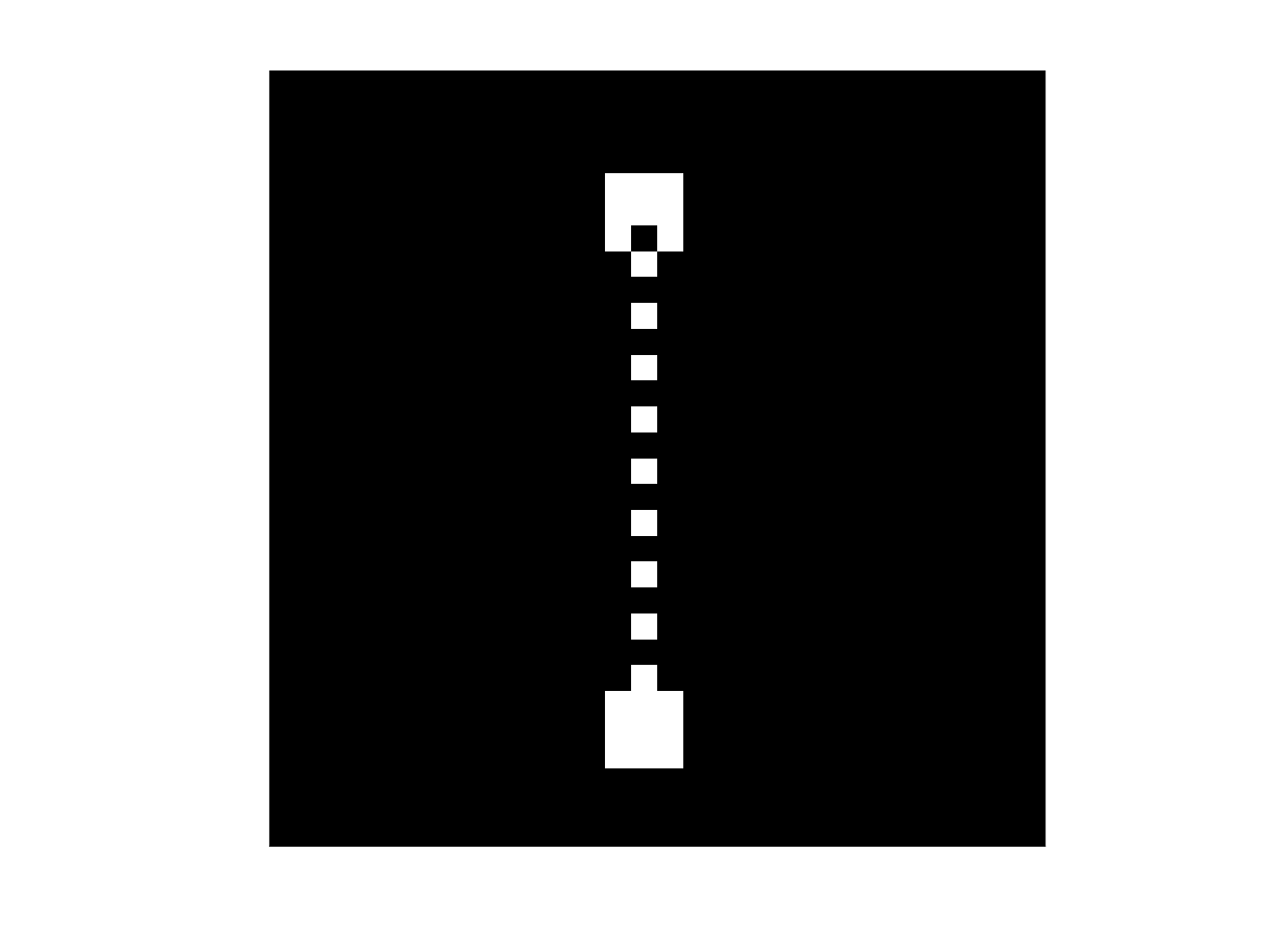}
	\caption{A wire carrying ``1'' (right) and a wire carrying ``2'' (left).}
	\label{w11}
\end{figure}

\begin{figure}[htb]
	\centering
	\psfrag{a}[c]{\color{yellow} $y$}
	\psfrag{b}[c]{\color{yellow} $x$}
	\psfrag{c}[c]{\color{yellow} $y$}
	\psfrag{d}[c]{\color{yellow} $z$}
	\includegraphics[width=0.5\textwidth]{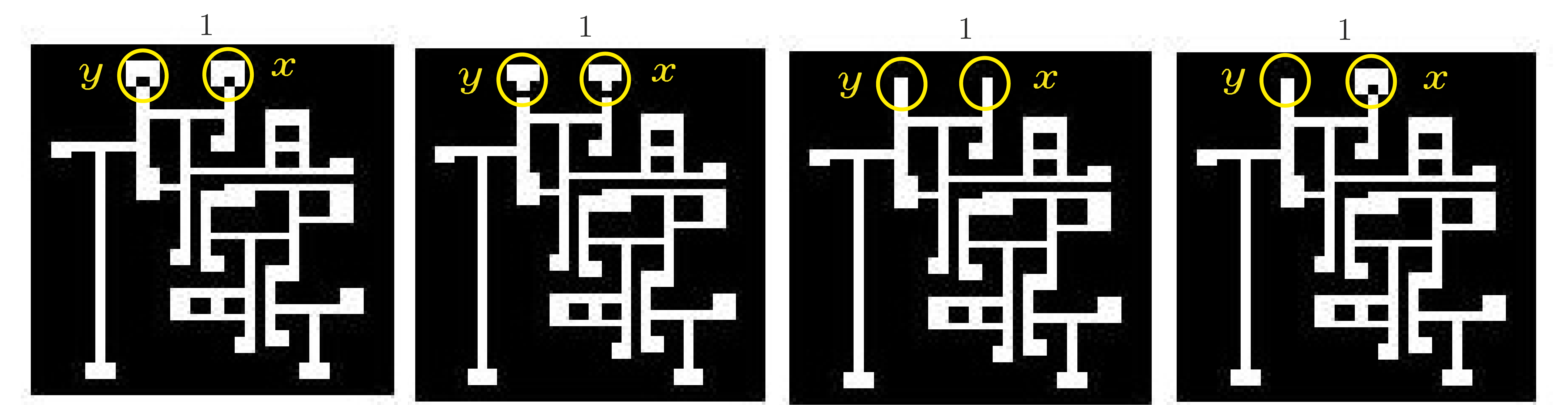} \\
	\includegraphics[width=0.5\textwidth]{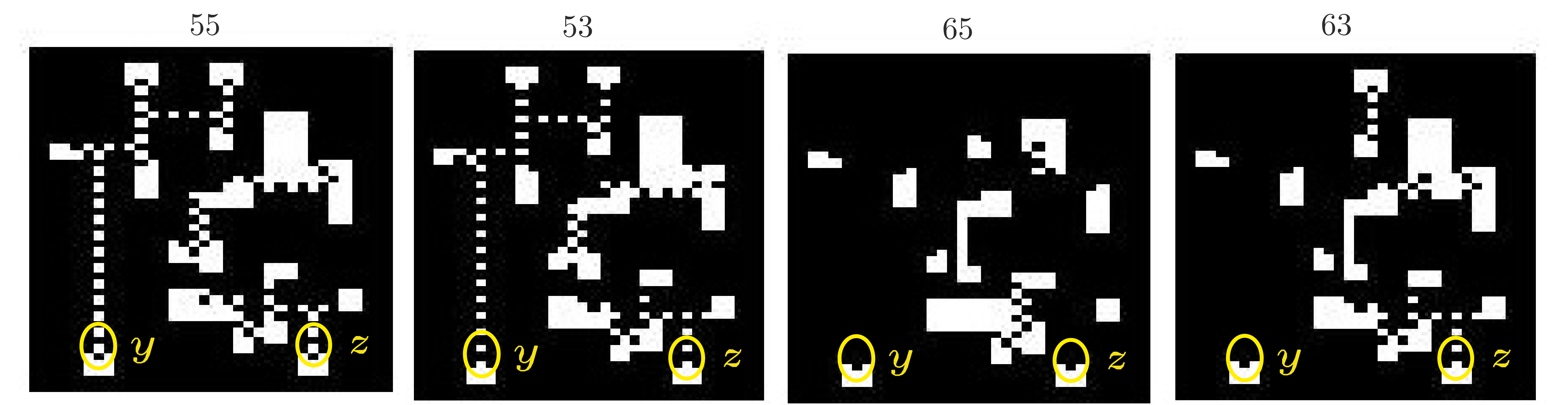}
	\caption{A 2-input-2-output trinary logic gate. From left to right the
		inputs are $(1,1)$, $(2,2)$, $(0,0)$, and $(0,1)$. The grid size is $43\times34$.}
	\label{crossing}
\end{figure}

\begin{figure}[htb]
	\centering
	\psfrag{a}[c]{\color{yellow} $y$}
	\psfrag{b}[c]{\color{yellow} $x$}
	\psfrag{c}[c]{\color{yellow} $y$}
	\psfrag{d}[c]{\color{yellow} $z$}
	\includegraphics[width=0.5\textwidth]{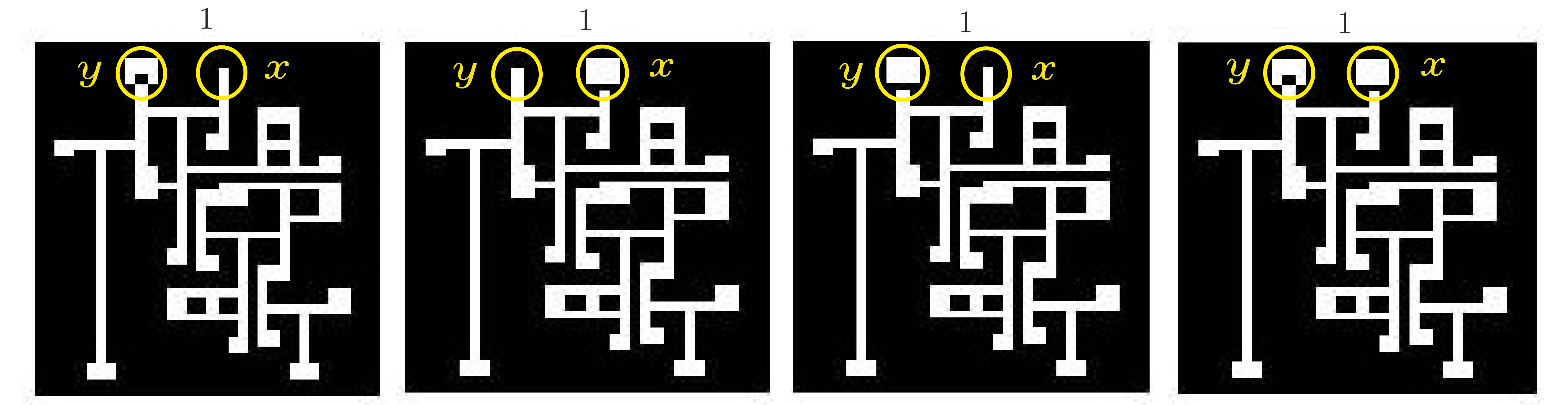} \\
	\includegraphics[width=0.5\textwidth]{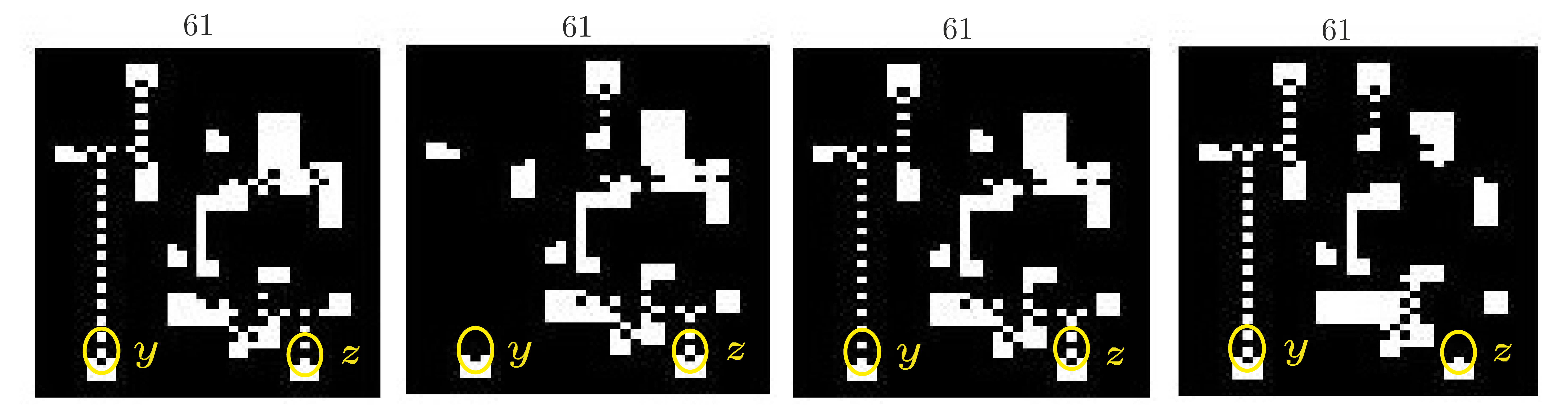}
	\caption{A 2-input-2-output trinary logic gate. From left to right the
		inputs are $(1,0)$, $(0,2)$, $(2,0)$, and $(1,2)$. The grid size is $43\times34$.}
	\label{crossing1}
\end{figure}

\begin{figure}[htb]
	\centering
	\psfrag{a}[c]{\color{yellow} $y$}
	\psfrag{b}[c]{\color{yellow} $x$}
	\psfrag{c}[c]{\color{yellow} $y$}
	\psfrag{d}[c]{\color{yellow} $z$}
	\includegraphics[width=0.2\textwidth]{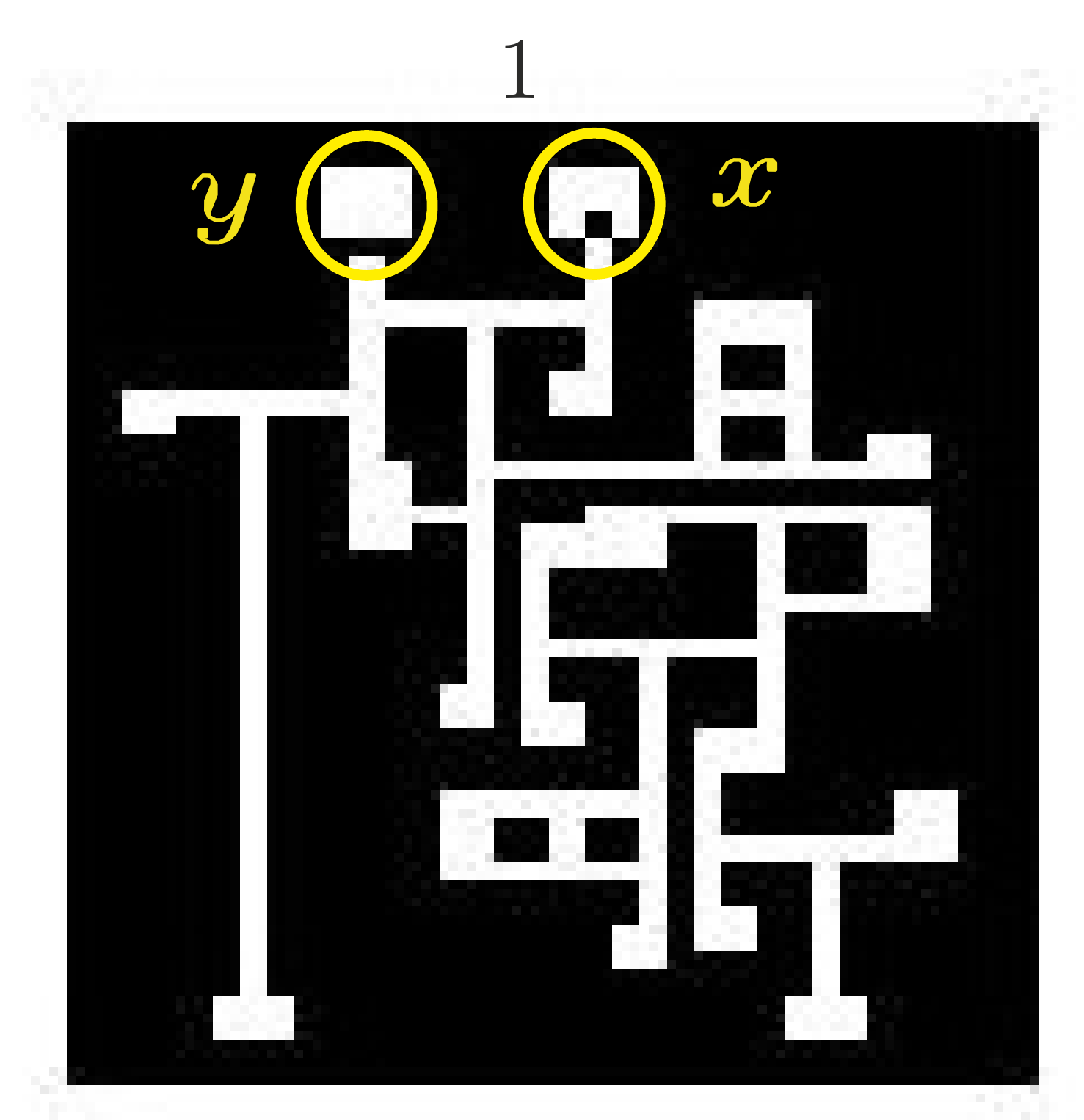} \\
	\includegraphics[width=0.2\textwidth]{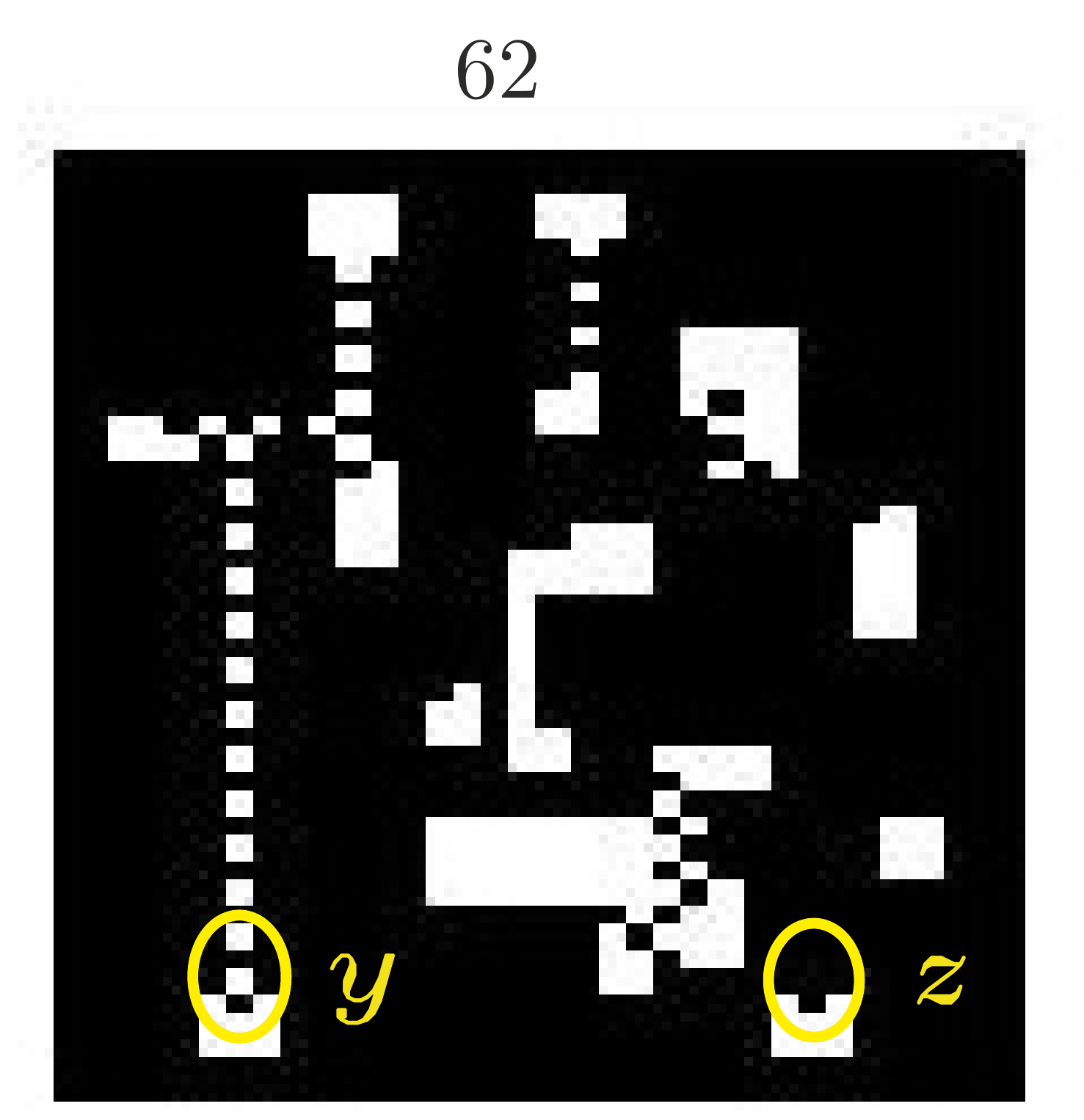}
	\caption{A 2-input-2-output trinary logic gate whose inputs are $(2,1)$. The grid size is $43\times34$.}
	\label{crossing2}
\end{figure}

In the underlying automaton the additional value ``2'' is represented
as ``1'' with a phase-shift. See Figure~\ref{w11}. The
realization of the 2-input-2-output crossing gate is shown for
the nine different input combinations in Figures~\ref{crossing},
\ref{crossing1}, and \ref{crossing2}. As before, the upper row in each
picture shows the grid at the beginning and the lower row shows its
state after a number of iterations of the automaton rule.

\subsection*{Gliders}

Gliders are animated entities that emerge in the grid during the
automaton evolution. In terms of computation such patterns are
instrumental for carrying information across the grid. The Life-based
Turing machine, for example, heavily relies on gliders to realize its logic and
memory parts~\cite{rendell2016game}.

For reasons mentioned in the introduction, we suspect that the
preceding cellular automaton cannot produce gliders. We were
able, however, to generate gliders with an automaton whose rule
permits at times the increase of a cell's Kolmogorov complexity.  One
cycle of this automaton is as follows. In the beginning of a cycle it
employs two rules to obtain the grid in the next time step:\\[0.5ex]

\noindent
{\bf Rule:} \emph{Nothing comes out of nothing -- do nothing to a
  (blank) cell whose Moore neighborhood vanishes.}\\[0.5ex]

\noindent
{\bf Rule:} \emph{A cell's value is changed
from $c$ to $1-c$ if the Kolmogorov complexity of its present Moore
neighborhood is larger with $1-c$ than with $c$.}\\[0.5ex]

\noindent
A single cycle of this automaton starts with a single iteration of
Algorithm~\ref{alg:the_alg2}. For the next few time steps the
automaton operates as described in
Algorithm~\ref{alg:the_alg1}, i.e. it employs the ``nowhere increasing''
complexity rule. It proceeds so until the pair of grids, the recent
one at an odd time step and the one two time steps back, are the
same. This cycle is repeated indefinitely. A pseudocode for this
automaton is given in Algorithm~\ref{alg:the_alg3}.

\begin{algorithm}
	\caption{Nowhere decreasing Kolmogorov complexity CA }
	\label{alg:the_alg2}
	\begin{algorithmic}
		\STATE Syntax: $c_{ij}(t+1) = \text{\bf CA}\uparrow(c_{ij}(t))$
		\STATE Input: $c_{ij}(t)$, $i=1,\ldots, N$, $j=1,\ldots, M$ (grid at
		time $t$)
		\STATE Output: $c_{ij}(t+1)$, $i=1,\ldots, N$, $j=1,\ldots, M$ (grid
		at time $t+1$)
		\FOR{$i = 2:N-1$}
		\FOR{$j = 2:M-1$}
		\STATE Let $A$ be the Moore neighborhood of $c_{ij}(t)$.
		\IF{$A$ does not zeros}
		\STATE Obtain $K_{ij}(t)$ using $A$ from the lookup table \cite{zenil2015two}.
		\STATE Flip the value of the middle cell in $A$ and similarly obtain
		$K'_{ij}(t)$.
		\IF{$K_{ij}(t) \geq K'_{ij}(t)$}
		\STATE $c_{ij}(t+1) = c_{ij}(t)$
		\ELSE
		\STATE $c_{ij}(t+1) = 1 - c_{ij}(t)$
		\ENDIF
		\ENDIF
		\ENDFOR
		\ENDFOR

	\end{algorithmic}
\end{algorithm}

\begin{algorithm}
	\caption{Alternating, nowhere decreasing - nowhere increasing, CA}
	\label{alg:the_alg3}
	\begin{algorithmic}
		\STATE Syntax: $c_{ij}(s) = \text{\bf CA}\updownarrow(c_{ij}(t))$
		\STATE Input: $c_{ij}(t)$, $i=1,\ldots, N$, $j=1,\ldots, M$ (grid at
		time $t$)
		\STATE Output: $c_{ij}(s)$, $i=1,\ldots, N$, $j=1,\ldots, M$ (grid
		at time $s$)
		
		\STATE $c_{ij}(t+1) = \text{\bf CA}\uparrow(c_{ij}(t))$
		
		\STATE $s = t+1$
		\WHILE{$\{ \exists i,j \; \;  c_{ij}(s) \neq c_{ij}(s-2) \}
			\vee \{s \text{ is even} \}$}
		\STATE $c_{ij}(s+1) = \text{\bf CA}\downarrow(c_{ij}(s))$
		\STATE $s=s+1$
		\ENDWHILE 
	\end{algorithmic}
\end{algorithm}

\begin{figure}[htb]
	\includegraphics[width=0.45\textwidth]{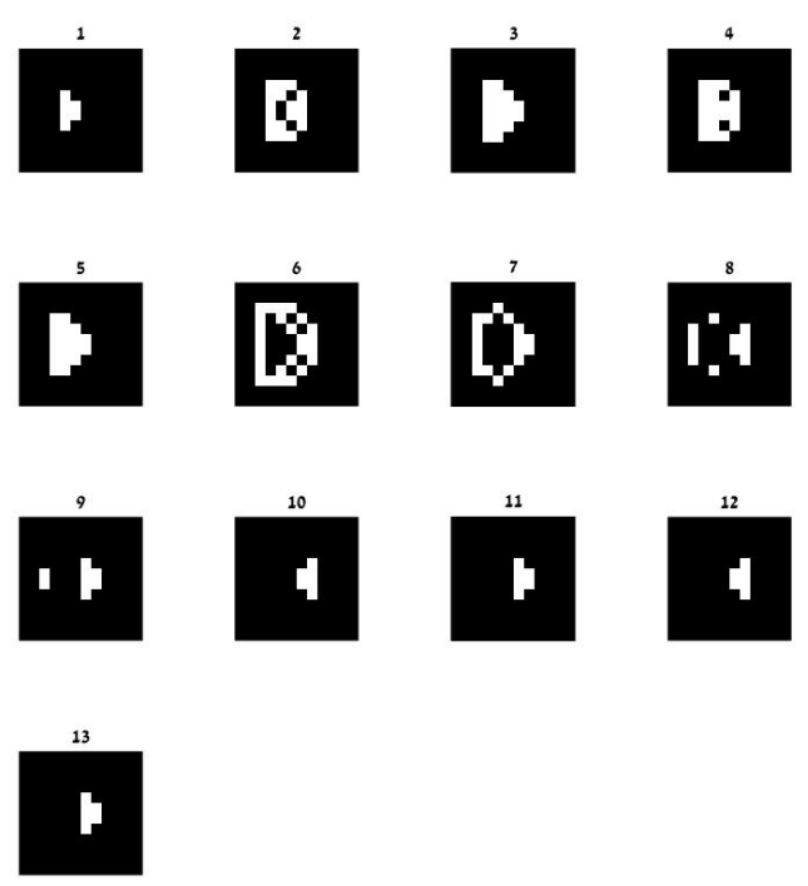}
	
	\caption{A two-cycle evolution of a glider using the cellular automaton in
		Algorithm~\ref{alg:the_alg3}. The first cycle starts with the grid
		numbered 1 and concludes with the grid numbered 5. The second cycle
		starts with grid 5 and concludes with grid 13. The grid size is $12\times12$.}
	\label{fig:glider}
\end{figure}

The basic construction of a glider and its evolution in the course of
two cycles of this automaton is shown in Figure~\ref{fig:glider}.

\section{Grid's average complexity}

\begin{figure}[htb]
	\psfrag{x}[c]{Time step}
	\psfrag{y}[c]{$\bar{K}$}
	\includegraphics[width=0.5\textwidth]{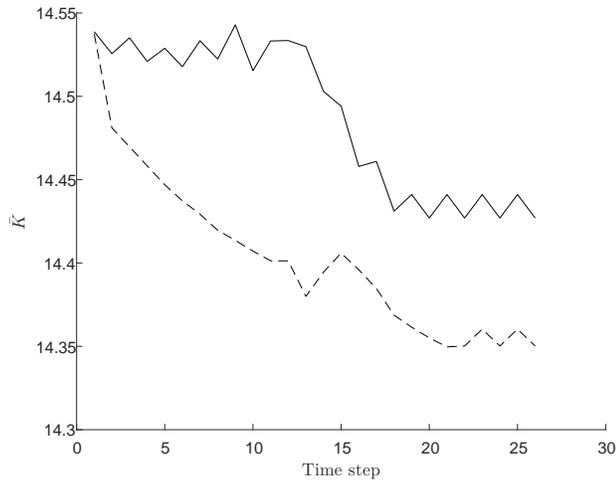}
	
	\caption{The average Kolmogorov complexity during the evolution of the
		NOT gate in Figure~\ref{not}. The solid and dashed lines correspond,
		respectively, to the inputs $1$ and $0$.}
	\label{kg}
\end{figure}

As neighborhoods overlap the average Kolmogorov complexity in the grid
may increase even in ``nowhere increasing'' mode. The average
Kolmogorov complexity of an $N \times M$ grid at time $t$ is defined
as
\begin{equation}\label{eq:kaverage}
\bar{K}(t) = (N-2)^{-1}(M-2)^{-1} \sum_{i=2}^{N-1} \sum_{j=2}^{M-1} K_{ij}(t).
\end{equation}
When evaluating this measure for the NOT gate in Figure~\ref{not} the
behavior in Figure~\ref{kg} is observed. The transition from the
initial to the final grid, in which complexity is lower on the
average, shows instances where the average complexity rises.

\begin{figure}[htb]
\centering
	\psfrag{a}[c]{\color{yellow} $x$}
	\psfrag{b}[c]{\color{yellow} $y$}
	\includegraphics[width=0.45\textwidth]{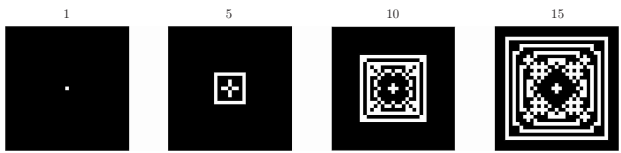}
	
	\caption{Expanding behavior of one source using the cellular automaton in
		Algorithm~\ref{alg:the_alg2}. The grid size is $30\times30$.}
	\label{bigb}
\end{figure}

\begin{figure}[htb]
	\psfrag{x}[c]{Time step}
	\psfrag{y}[c]{$\bar{K}$}
	\includegraphics[width=0.45\textwidth]{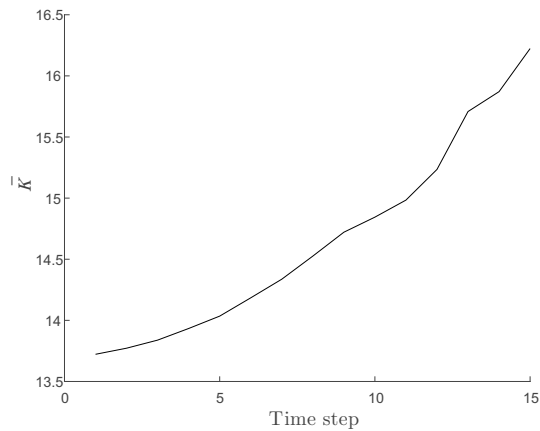}
	
	\caption{The average Kolmogorov complexity during the evolution of the
		the automaton in Figure~\ref{bigb}.}
	\label{kg1}
\end{figure}

Evaluating this measure for the automaton in
Algorithm~\ref{alg:the_alg2} results in the typical behavior shown in
Figure~\ref{bigb}. As one expects, its average complexity in
Figure~\ref{kg1} tends to increase in time.

\section{Conclusion}\label{sec:Conclusion}

This work is an attempt to address the questions raised in the beginning
of this paper. The answer we offer is only partial. One may
wonder whether any computable function can be computed by a similar
``nowhere increasing'' cellular automaton, or in other words, whether such an
automaton is Turing-complete. For one reason we think it isn't. During its
evolution the initial grid on which the logical gate is encoded
self-destructs. Therefore, outputs cannot be reused as inputs to the
same logical gate. Although not proven, we suspect that this behavior
hinders the construction of a memory device and thus also of a
Turing-equivalent model of computation.

But the concept of using a measure of complexity to evolve is
multifaceted. We have shown that a cellular automaton whose rule
permits at times the increase of the cell's neighborhood complexity
can produce gliders. The lesson learned from Life is that gliders may
become the basic ingredients in any computation and so perhaps this
automaton also is Turing-complete.


As a final remark, we have used a particular measure of
complexity. Other measures may similarly be used to evolve
cellular automatons with different computational capabilities and
behavior.



\end{document}